\documentclass[aps,twocolumn,superscriptaddress,showpacs,floatfix]{revtex4}
\usepackage[bf]{caption}
\usepackage{graphicx}
\usepackage{amsmath}
\usepackage{wrapfig}

\begin{document}

\title{The role of Coulomb interaction in fragmentation}
\author{M.J. Ison}
\affiliation{ Departamento de F\'{i}sica, Facultad de Ciencias Exactas y Naturales, Universidad de Buenos Aires,
 Pabell\'on $I$, Ciudad Universitaria, Nu\~{n}ez, $1428$,\\
 Buenos Aires, Argentina.}
\email{mison@df.uba.ar}
\author{C.O. Dorso}
\affiliation{ Departamento de F\'{i}sica, Facultad de Ciencias Exactas y Naturales, Universidad de Buenos Aires,
 Pabell\'on $I$, Ciudad Universitaria, Nu\~{n}ez, $1428$,\\
 Buenos Aires, Argentina.}

\date{\today}

\begin{abstract}
We examine the impact of adding a Coulomb term to a constrained system of $147$ particles interacting via a 
Lennard-Jones ($LJ$) potential, finding that the inclusion of the coulombic interaction produces a shift, 
but no qualitative changes in the thermodynamical properties of the system. We also performed the systems 
characterization from a morphological point of view.

\end{abstract}


\maketitle

The analysis of the caloric curves ($CC$) of small systems is attracting the attention of physicists in 
different areas. One of the most active ones is precisely nuclear physics because of the possibility of the 
occurrence of phase transitions in multifragmentation experiments. A wealth of theoretical and experimental 
work has flourished in recent years with, in some cases contradictory results. In particular a debate has 
recently arisen regarding the effect of Coulomb interaction on the properties of the $CC$. Whereas some 
authors {references \cite{chomazprl,gulmiarxiv}} claim that there exist a loop in the $CC$ denoting a negative 
thermal response function for systems as large as $200$ particles it has also been claimed that an upper limit 
in the mass exist ($A=60$) \cite{moretto,morettoarxiv} above which no such a loop can exist. In Ref.\cite{raduta}
 it is stated that the presence of Coulomb forbids the phase transition for systems as large as $A=200, Z=82$ 
(a proton fraction of $0.4$). On the other hand in Ref.\cite{das} they propose a model in which $c_v$ is 
never negative.  

In this communication we will show that, when dealing with finite constrained systems interacting via a 
Lennard-Jones $6-12$ potential which has been shown to display negative values of the specific heat [or more 
generally speaking, the thermal response function ($TRF$)] for low values of the density \cite{noneqfrag}, the 
addition of the Coulomb interaction term preserve the shape of the caloric curve and gives rise to a 
displacement of the location of the loop together with its flattening. These results are consistent with 
lattice models calculations 
including Coulomb interaction \cite{richert}. It is worth to mention that the relevance of the analysis 
of $LJ$ systems relies on the fact that the $EOS$ of the $LJ$ and the one assumed to describe nuclear matter 
are quite similar \cite{Pandha,cmdPandha}. 

The system under study is composed by a gas of $147$ particles confined in a spherical box, defined by the 
Hamiltonian $H=K+V_{LJ}+V_{coulomb}+V_{walls}$, where $K$ is the kinetic energy and the short-range term of the 
interaction potential is given by $V_{LJ}=\Sigma v(r_{ij})$ with $v(r_{ij})=4 \epsilon %
[ \frac{(\sigma}{r_{ij})}^{12}-\frac{(\sigma}{r_{ij})}^{6}]$ if $(0<r<r_c)$ and $0$ otherwise. In this work we 
took a cut-off radius $r_c=3.0\sigma$, and adopted adimensional units for the energy, length and time such that 
$\epsilon=\sigma=1$, $t_{0}=\sqrt{\sigma^{2}m/48\epsilon}$.

The role of $V_{walls}$ is to constrain the particles inside an spherical container. The considered external 
potential behaves like $V_{wall}\sim (r-r_{wall})^{-12}$ with a cut-off distance $r_{cut}=1\sigma$, where it 
smoothly became zero along with its first derivative. 

The addition of the coulombic interaction $V_{coulomb}$ in the hamiltonian has been done with two different 
approaches that will be henceforth referenced as the homogeneous and the inhomogeneous one.

The homogeneous case, which could be useful in the study of metallic clusters and has already been introduced
for the study of nuclear fragmentation \cite{Pandha}, the coulombic force is included between all particles
 of the fluid, $V_{coulomb}=q^2/r_{ij}$, with $q^2=0.055$, wich was obtained in \cite{Pandha} by comparing the 
binding energy formula for argon balls with the nuclear drop mass formula assuming the proton fraction to be 
$0.4$ (roughly speaking, it has been adjusted so that liquid argon drops having $\sim 300$ atoms are unstable 
under fission).   

In the second case, i.e. the inhomogeneous case, the Coulomb term is only present between a subset ($Z$) of 
the ($A$) particles which will be renamed as "protons" (the other $A-Z$ will therefore called "neutrons"). 
Taking the same proton fraction as in the previous case ($\frac{Z}{A}=0.4$), we adjusted the strength of the 
Coulomb term so that we have the same total energy ($E=K+V_{LJ}+V_{coulomb}+V_{walls}$) for the same 
configuration. The value of $q^2$ turned out to be $q^2=0.238$. This last approach mimics the nuclear scenario
 in a more realistic way.

For both systems we have performed extensive molecular dynamics simulations following the same approach as 
described in \cite{noneqfrag}.  

\begin{figure}
\centering
\includegraphics[width=6cm,clip=]{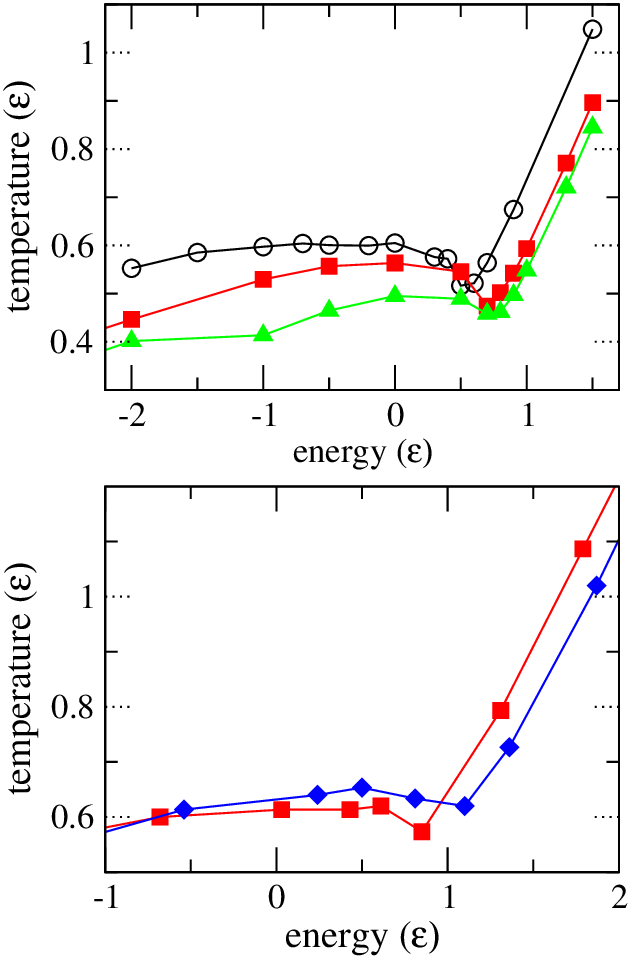}%
\caption{Top:Caloric curves for $\rho=0.01 \sigma^{-3}$. Pure LJ (empty circles), inhomogeneous 
(filled squares) and homogeneous (filled triangles) systems. Bottom: Caloric curves for an inhomogeneous 
system with $\rho=0.015 \sigma^{-3}$. $A=147$, $Z=62$ (squares) and $A=294$, $Z=124$ (diamonds). \label{cc}}
 \end{figure}

We will first focus on the thermodynamics of these systems at a very low densities $(\rho < 0.025 \sigma^{-3})$.
 For these densities a pure $LJ$ system displays a well defined loop in its $CC$ as a direct consequence of 
the ability of the system to form well defined fragments in configuration space \cite{noneqfrag}. In the upper 
panel of Fig.~\ref{cc} we show the caloric curves for a pure $LJ$ (empty circles), $LJ+Coulomb$ inhomogeneous 
case (full triangles) and homogeneous case (full squares) for $\rho=0.01 \sigma^{-3}$. 
It can be seen that for all the three cases a clear loop is displayed. If we restrict our analysis to the 
systems with Coulomb interaction we notice that the location of the minima of the loop is shifted with respect 
to the $LJ$ case. This shift is of the order of $0.3\epsilon$, which corresponds to the mean coulombic energy 
of the system. This can be easily verified by plotting the temperature as a function of $K+V_{LJ}$, in which 
case the shift dissapears denoting that the partition of the total potential energy is made into two terms: 
The LJ potential energy which is extremely sensitive to the presence of inner surfaces, and the coulomb term, 
which is, on the contrary, very insensitive due to its long range.  

In the lower panel of Fig.~\ref{cc} we illustrate the fact that the presence of the loop is quite resilient to 
an increase of the systems size. In particular we show the $CC$ for $\rho=0.023 \sigma^{-3}$ and $A=294$ for the pure $LJ$ system and the inhomogeneous case (with $Z=124$). It can be seen that both the loop and the shift 
remain present. 

\begin{figure}
\centering
\includegraphics[width=6cm,clip=]{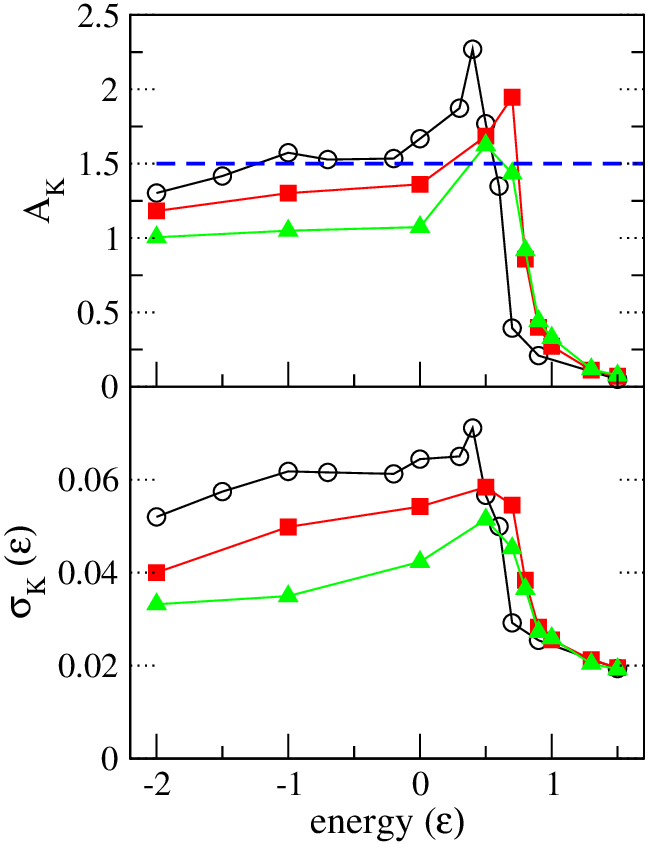}%
\caption{Top:Relative fluctuations of kinetic energy for $\rho=0.01 \sigma^{-3}$. Pure LJ (empty circles), 
inhomogeneous (filled squares) and homogeneous (filled triangles) systems. Bottom: Standard deviation of 
kinetic energy per particle for the same cases.\label{fluct}}
 \end{figure}

In order to gain further insight into the properties of such a system we studied the second moments of the 
distribution of kinetic energies, namely the standard deviation of the kinetic energy per particle and the 
relative kinetic energy fluctuations, defined as      

\begin{equation}
A_K=N \frac{\sigma^2_K}{T^2}
\label{eqA}
\end{equation}

where $N$ is the number of particles, $\sigma_K$ the standard deviation of the kinetic energy 
per particle and $T$ the temperature of the system. Since kinetic energy fluctuations and the specific heat 
are related by \cite{lebowitz} 

\begin{equation}
 N <\sigma_K^2>_E={\frac{3}{{2 \beta^2}}} (1 - {\frac{3 }{{2 C}}%
})  \label{eqFluc}
\end{equation}

Negative values of the specific heat should be expected whenever $A_K$ exceeds the canonical $A_K(can)=1.5$ 
\cite{gulmi_fluc, noneqfrag}.

In Fig.~\ref{fluct} we show these two quantities, $A_K$ (upper panel) and $\sigma_K$ (lower panel). Once again 
the presence of the loop in the $CC$ is correlated to the peaks in the values of $A_K$ and $\sigma_K$. 
Moreover these peaks are above the canonical value (dashed line of Fig.~\ref{fluct}) thus denoting a negative 
value of the $TRF$. It is also interesting to notice that for the homogeneous case the size of the peak is 
strongly reduced, which is in accordance with the fact that the loop in the corresponding $CC$ is shallower. 
On the other hand for the inhomogeneous case this effect is much weaker.

\begin{figure}
\includegraphics[width=6cm,clip=]{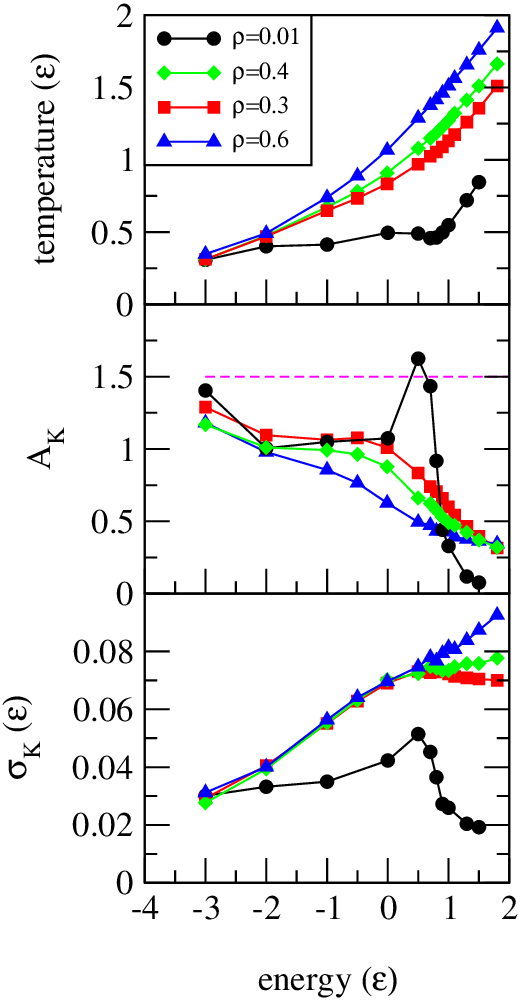}%
\caption{Caloric curve for the homogeneous $LJ+coulomb$ case for different densities (upper panel). 
Relative fluctuation of the kinetic energy (medium panel) (notice that the peak of $A_K$ exceeds the $1.5$ 
canonical value for $\rho=0.01 \sigma^{-3}$). Standard deviation of the  kinetic energy per particle (bottom).
\label{thermodens}}
\end{figure}

Following the scheme used in reference \cite{phasediagram} we now explore the effect of increasing the density 
in the above presented quantities. The results are summarized in Fig.~\ref{thermodens} in wich we show the 
temperature (upper panel), $A_K$ (middle panel) and $\sigma_K$ (lower panel) for different densities (see 
figure caption for details). In the case of the temperature it displays a transition from a $CC$ with a loop 
($\rho=0.01 \sigma^{-3}$) to a monotously increasing function for the highest density considered 
($\rho=0.60 \sigma^{-3}$). For intermediate density values the $CC$ displays a change of slope. These features 
are in complete agreement with the already calculated behavior of the $CC$ for the pure $LJ$ case 
\cite{phasediagram}. The behavior displayed by $A_K$ and $\sigma_K$ (middle and lower pannels respectively) 
is also similar to those found for the pure $LJ$ case. These results confirm the already pointed "triviality" 
of the inclusion of Coulomb interaction for small constrained systems. It is worth to mention at this point 
that the behavior of the unnormalized fluctuation of the kinetic energy $\sigma_K$ changes qualitatively from 
displaying a loop to an increasing function of the energy at a density between $\rho=0.3 \sigma^{-3}$ and 
$\rho=0.4 \sigma^{-3}$ which has been pointed out \cite{campicoulomb} to be the critical density for an 
homogeneous $LJ+Coulomb$ system.      

\begin{figure}
\center
\includegraphics[width=9cm]{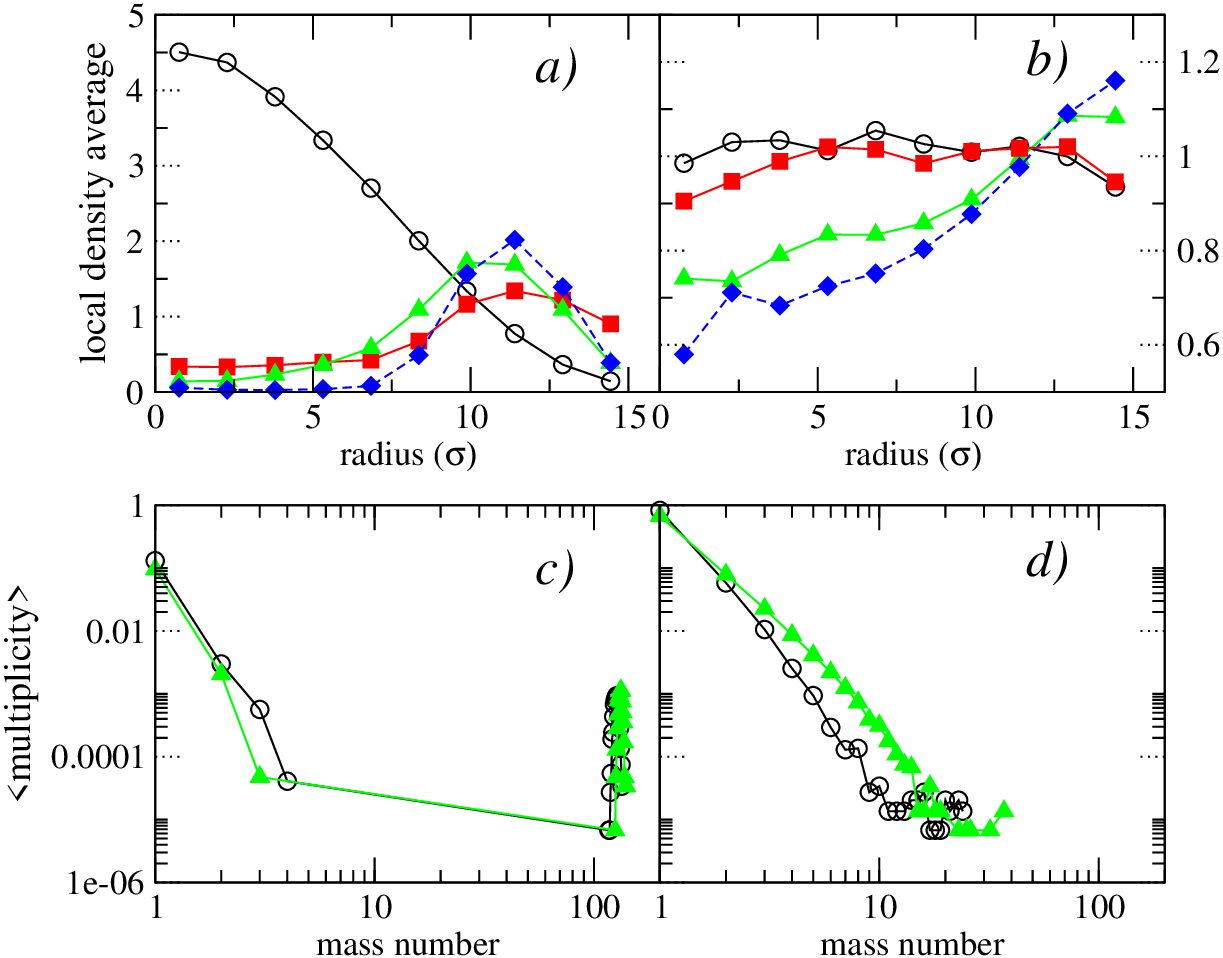}
\caption{ Panels ($a$) and ($b$): Local density profiles averages ({\it LDPA}) for a liquid-like configuration 
($a$) and a gas-like configuration ($b$). $LJ$ {\it LDPA} (empty circles), homogeneous $LJ+Coulomb$ {\it LDPA} 
(filled triangles), proton {\it LDPA} (filled diamonds), and neutron {\it LDPA} (filled squares). Panels ($c$) 
and ($d$): Mass distributions for the $LJ$ (empty circles) and homogeneous $LJ+Coulomb$ (filled triangles) 
cases.}\label{spectra}
\end{figure}

So far the presence of Coulomb interaction has not modified any of the "thermodynamic" properties of the system.
 We now turn to the analysis of its morphological properties. We first study the density profiles for the three 
cases studied in this work. For this purpose we divide our system in concentrical equally spaced shells and 
calculate the normalized average populations of each shell. We have performed this analysis for the low 
density case for two relevant energies, namely $E=-2.0 \epsilon$ and $E=1.0 \epsilon$ for the $LJ+Coulomb$ 
systems and $E=-2.7 \epsilon$ and $E=0.7 \epsilon$ for the $LJ$ case (panels ($a$) and ($b$) in 
Fig.~\ref{spectra}). To properly understand this
 behavior we have performed a fragment analysis of the system. We have defined fragments according to the 
following definition (see \cite{hill,ale97b}): Given a set of particles $i, j,..., k$, it belongs to the same 
cluster $C_i$ if:
\begin{equation}\forall \, i \, \epsilon \, C_i \:,\: \exists \, j \,
\epsilon \, C_i  \, /
\,  e_{ij} \leq V_{max}
\end{equation}
where $e_{ij} = V(r_{ij}) + ({\bf p}_i - {\bf p}_j)^2 / 4 \mu$, $\mu$ is the reduced mass of the pair 
$\{i,j\}$, and $V_{max}$ is an upper limit for the potential energy ($0$ in the $LJ$ case).

This definition takes into account in an approximate way the relative momentum of the particles. The 
corresponding mass distributions are displayed in panels ($c$), ($d$) (for the same energies depicted as 
before). 
For the sake of clarity we only show the fragment mass spectra for the pure $LJ$ and the homogeneous case. 
We can see that for pure $LJ$ and homogeneous case a big fragment is formed together with small aggregates 
giving rise to a U-shape distribution so the difference in the density profiles (circles and triangles in 
panels ($a$), ($b$) for $LJ$, $LJ+Coulomb$ respectively) comes from the fact that the big fragment for the 
pure $LJ$ case travels along the whole volume remaining mainly in the center of the container whereas for the 
homogeneous case it remains close to the walls due to Coulomb repulsion with the rest of the system. For the 
high energy case the system is mainly composed of small fragments, in both analized cases, but while for the 
pure $LJ$ case particles are homogeneously distributed inside the volume, for the charged system the effect 
of repulsion gives rise to a higher density close to the walls. 

Turning our attention to the inhomogeneous case, it is interesting to notice the differences between the 
proton and neutron density profiles (squares and diamonds in panels $a)$ and $b)$ of Fig.~\ref{spectra} 
respectively). For the low energy case no major differences are found, which is related to the fact that the 
structure of the big clusters is composed by both protons and neutrons. However, the scenario changes 
completely when looking at the system at high energies (panel ($b$)). Since the system is now composed mainly 
of small clusters the density profile of the protons is quite similar to the homogeneous case, with higher 
values close to the walls due to Coulomb repulsion, but the neutron density profile now clearly resembles 
the $LJ$ profile i.e. an homogeneous distribution inside the container.     

To sum up, after analyzing both thermodynamical (caloric curves, kinetic energy fluctuations) and 
morphological properties (fragment mass distributions, density profiles) of systems with and without Coulomb 
interaction we came up with the result that phase transitions in small systems, with masses of the order of 
those relevant for the nuclear case, are not substantially modified by the presence of such long range forces.
 Nevertheless a slight reduction of the signals takes place specially for the homogeneous case, which could 
be useful in the study of metallic clusters, where negative specific heats have been measured experimentally
 \cite{schmidt}. 

We thank Ariel Chernomoretz for useful discussions and Francesca Gulminelli for a critical reading of the 
manuscript. This work was partially supported by the University of Buenos Aires via grant X139.


\bibliographystyle{ieeetr}
\bibliography{BRcoulomb.bib}

\begin{thebibliography}{10}

\bibitem{chomazprl}
P.~Chomaz, V.~Duflot, and F.~Gulminelli {\em Phys. Rev. Lett.}, vol.~85,
  p.~3587, 2000.

\bibitem{gulmiarxiv}
F.~Gulminelli and P.~Chomaz, 2003.
\newblock arXiv:nucl-th/0304058.

\bibitem{moretto}
L.~G. Moretto, J.~B. Elliott, L.~Phair, and G.~J. Wozniak {\em Phys. Rev. C},
  vol.~66, p.~041601, 2002.

\bibitem{morettoarxiv}
L.~G. Moretto, J.~B. Elliott, and L.~Phair, 2003.
\newblock arXiv:nucl-th/0307102.

\bibitem{raduta}
A.~H. Raduta and A.~R. Raduta {\em Phys. Rev. Lett.}, vol.~87, p.~202701, 2001.

\bibitem{das}
C.~B. Das, S.~D. Gupta, and A.~Z. Mekjian {\em Phys. Rev. C}, vol.~68,
  p.~014607, 2003.

\bibitem{noneqfrag}
A.~Chernomoretz, M.~Ison, S.~Ort\'iz, and C.~Dorso {\em Phys. Rev. C}, vol.~64,
  p.~024606, 2001.

\bibitem{richert}
J.~M. Carmona, J.~Richert, and P.~Wagner {\em Eur. Phys. J. A}, vol.~11, p.~87,
  2001.

\bibitem{Pandha}
R.~J. Lenk and V.~R. Pandharipande {\em Phys. Rev. C}, vol.~34, p.~177, 1986.

\bibitem{cmdPandha}
T.~J. Schlagel and V.~R. Pandharipande {\em Phys. Rev. C}, vol.~36, p.~162,
  1987.

\bibitem{lebowitz}
J.~L. Lebowitz, J.~K. Percus, and L.~Verlet {\em Phys. Rev}, vol.~153, p.~250,
  1967.

\bibitem{gulmi_fluc}
F.~Gulminelli, P.~Chomaz, and V.~Duflot {\em Europhys. Lett.}, vol.~50, p.~434,
  2000.

\bibitem{phasediagram}
A.~Chernomoretz, P.~Balenzuela, and C.~Dorso {\em Nucl. Phys. A}, vol.~723,
  p.~229, 2003.

\bibitem{campicoulomb}
X.~Campi, H.~Krivine, E.~Plagnol, and N.~Sator {\em Phys. Rev. C}, vol.~67,
  p.~044610, 2003.

\bibitem{hill}
T.~Hill, {\em Thermodynamics of small systems}.
\newblock Dover, 1994.

\bibitem{ale97b}
A.~Strachan and C.~O. Dorso {\em Phys. Rev. C}, vol.~56, p.~995, 1997.

\bibitem{schmidt}
M.~Schmidt, R.~Kusche, T.~Hippler, J.~Donges, W.~Kronmüller, B.~von
  Issendorff, and H.~Haberland {\em Phys. Rev. Lett.}, vol.~86, p.~1191, 2001.

\end{thebibliography}

\end{document}